\documentclass[
 reprint,
superscriptaddress,
amsmath,amssymb, 
aps,
pre,
]{revtex4-2}

\usepackage{xcolor}
\definecolor{refblue}{RGB}{35,75,120}
\usepackage[
    colorlinks=true,
    linkcolor=refblue,
    citecolor=refblue,
    urlcolor=refblue,
    breaklinks=true
]{hyperref}
\usepackage[normalem]{ulem}
\usepackage[english]{babel}
\usepackage{dsfont}
\usepackage{mathrsfs}
\usepackage{bm}
\usepackage{graphicx}
\usepackage{dcolumn}
\usepackage{amsmath}
\usepackage{placeins}
\usepackage{comment}
\usepackage[bottom]{footmisc}
\usepackage{physics}
\usepackage{listings}
\usepackage{tabularx}
\usepackage{makecell}
\usepackage{bbold}
\usepackage{booktabs}
\usepackage{caption}
\newcommand{\RN}[1]{%
  \textup{\uppercase\expandafter{\romannumeral#1}}%
}

\begin{document}

\title{Physical Mechanism of Vacuole Formation in Liquid Droplets}

\author{Pranay Jaiswal}
\thanks{These authors contributed equally to this work.}
\affiliation{Faculty of Mathematics, Natural Sciences, and Engineering: Institute of Physics, University of Augsburg, Universit\"atsstra{\ss}e\ 1, 86159 Augsburg, Germany}

\author{Ivar S. Haugerud}
\thanks{These authors contributed equally to this work.}
\affiliation{Faculty of Mathematics, Natural Sciences, and Engineering: Institute of Physics, University of Augsburg, Universit\"atsstra{\ss}e\ 1, 86159 Augsburg, Germany}
\affiliation{
Department of Mathematics, Faculty of Mathematics and Natural Sciences, University of Oslo, Oslo 0315, Norway}

\author{William Verstraeten}
\affiliation{Heidelberg University, Center for
Molecular Biology of Heidelberg University (ZMBH), Berliner Str. 45, 69120 Heidelberg, Germany}

\author{Kerstin Göpfrich}
\affiliation{Heidelberg University, Center for
Molecular Biology of Heidelberg University (ZMBH), Berliner Str. 45, 69120 Heidelberg, Germany}
\affiliation{Max Planck School Matter to Life, Jahnstr. 29, 69120 Heidelberg}

\author{Job Boekhoven}
\affiliation{School of Natural Sciences, Department of Bioscience, Technical University of Munich, Lichtenbergstraße 4, 85748 Garching, Germany}

\author{Christoph A. Weber}
\email{christoph.weber@physik.uni-augsburg.de}
\affiliation{Faculty of Mathematics, Natural Sciences, and Engineering: Institute of Physics, University of Augsburg, Universit\"atsstra{\ss}e\ 1, 86159 Augsburg, Germany}

\begin{abstract}
Vacuoles have been observed in liquid droplets across variety of experimental systems, ranging from biomolecular condensates composed of proteins and RNA, to synthetic coacervates formed by charged polymers or synthetic nanostars. 
These vacuoles are long-lived domains depleted of droplet material, and their formation is puzzling because the associated increase in interfacial area is thermodynamically unfavorable. 
Using theory, we show that vacuoles form through a generic mechanism: a local spinodal instability within the droplet. 
We demonstrate this mechanism in several experimentally relevant scenarios, including temperature quenches and droplets coupled to chemical processes occurring either inside or outside the droplet. 
Using non-equilibrium thermodynamics, we develop a theoretical framework that identifies the physicochemical conditions controlling whether vacuoles form and how big vacuoles can become. Our results suggest molecular designs and chemical pathways that promote vacuolation, enabling multi-compartment formation with engineered functions such as enhanced surface catalysis and compartment fission.
\end{abstract}
\maketitle

Liquid droplets are shaped by a simple thermodynamic imperative: interfacial area is costly~\cite{RowlinsonWidom1982,deGennes2004}. 
Surface tension drives isolated droplets toward spheres and many-droplet systems toward coarsening by fusion~\cite{Siggia1979,BinderStauffer1974}, or Ostwald ripening~\cite{LifshitzSlyozov1961,Wagner1961}.
These processes reduce the total interface and thereby lower the system free energy. 
The reverse process -- the spontaneous creation of new interface inside a droplet -- is therefore strongly disfavored.

However, non-spherical liquid droplets with liquid inclusions have been observed across a wide range of experimental systems.
These inclusions are droplet-material-poor domains inside liquid droplets with droplet material concentration similar to the outside phase. 
They are often referred to as \textit{vacuoles} in the condensate literature, and represent one form of higher-order internal architecture in biomolecular condensates~\cite{Fare2021,Lyon2021}.  
Vacuoles in liquid droplets were observed \textit{in vivo} in protein condensates~\cite{SchmidtRohatgi2016} and \textit{in vitro} in biomolecular condensates governed by multivalent protein–RNA interactions~\cite{Boeynaems2019,Alshareedah2020}, or charge-mediated protein coacervation~\cite{Li2025, modi2026transient}. 
Vacuoles were also reported in synthetic or programmable macromolecular droplets, including electrostatic polyelectrolyte coacervates~\cite{Moreau2020,Yewdall2021, Bergmann2023, modi2026transient} and sequence-programmed DNA/RNA nanostar droplets~\cite{Saleh2023,verstraeten2025genetic}.
Vacuole formation was seen after very different perturbations, e.g., when changing the osmotic condition, composition, temperature or pH~\cite{Moreau2020,Alshareedah2020, Erkamp2023, modi2026transient},  and due to chemical reactions, such as enzymatic processes in DNA/RNA droplets~\cite{Saleh2023, verstraeten2025genetic, giessler2026growth, modi2026transient}, and fuel-driven reactions~\cite{Bartolucci2021, Bergmann2023, bauermann2023formation}.

Different microscopic mechanisms for vacuole growth were proposed.
In DNA nanostar droplets, internal restriction-enzyme cleavage of the droplet material produces DNA fragments that accumulate in the vacuole and raise its osmotic pressure; vacuole growth was therefore modeled as osmotic swelling against the Laplace pressure of the vacuole-droplet interface~\cite{Saleh2023}. 
In surfactant(DDAB)-coated complex coacervate droplets, by contrast, salt addition induced water-rich vacuoles together with an increase in droplet diameter, and the authors associated this behavior with the DDAB-coated interface and the double-chain architecture of DDAB, rather than with enzymatic production of internal osmolytes~\cite{Yin2023}. 
In liquid-crystalline, polysaccharide-surfactant coacervates, amylase-mediated hydrolysis triggered vacuolization and swelling into coacervate vesicles; here, the authors attributed the transformation to hydrolysis-induced weakening of multivalent electrostatic complexation, loss of liquid-crystalline order, and increased internal osmotic pressure~\cite{Jia2025}. 
Despite the microscopic differences, these systems have in common that a stimulus (cleavage, salt, hydrolysis) perturb the properties or composition of the dense phase.

The formation of vacuoles in liquid condensates remains less well understood. 
RNA addition to RNA-protein condensates was reported to produce ``hollow, vesicle-like condensates'' through phase separation of anisotropic protein-RNA complexes~\cite{Alshareedah2020}. 
However, interpreting these hollow structures as membrane-like is problematic: the authors' MD simulations show nanometer-scale vesicle-like assemblies, whereas the experimental vacuoles are micrometer-sized. This scale mismatch supports ``hollow condensate'' formation, but not  membrane formation as a mechanism.
More recently,  studies that quantitatively compared experiments and theory on RNA-nanostars~\cite{verstraeten2025genetic} and fuel-driven reaction cycles~\cite{Bergmann2023}, showed that a mesoscopic liquid droplet can transform into a liquid vacuole when the center concentration undergoes a diffusive instability. 
The frequent and robust observations of vacuoles in cells~\cite{SchmidtRohatgi2016} and various \textit{in vitro} biological~\cite{Boeynaems2019,Alshareedah2020,Li2025, modi2026transient} and synthetic systems~\cite{Moreau2020,Yewdall2021,modi2026transient, Saleh2023,verstraeten2025genetic,
Erkamp2023, Bartolucci2021, giessler2026growth, Bergmann2023, bauermann2023formation} pose the question of a common dynamical mechanism for vacuole formation and growth across these chemically very different systems subject to different perturbations.

Using theory, we show that vacuoles in liquid droplets can form across different systems, including binary and ternary systems, subject to various perturbations, such as temperature quenches and chemical reactions occurring in different phases.
Strikingly, vacuole formation in all such systems occurs through a common mechanism: a local diffusive, spinodal instability within the droplet.
We show how perturbations can give rise to concentration gradients within the droplet, where the concentration passes through the spinodal concentrations in the thermodynamic phase diagram. 
Specifically, if the perturbation is fast compared to diffusive transport within the droplet, or the droplet is big enough for fixed perturbation speed, the concentration response within the droplet lags behind.
The result is that the compositions within the droplet  undergo a spinodal instability locally. 
This diffusive instability forms one or multiple new interfaces that enclose domains with a molecular composition similar to that of the phase outside the droplet:  vacuoles are born. 
For chemical processes that perturb droplet composition and generate concentration gradients, we find that vacuoles form most readily when droplet material is degraded inside the droplet. 
We further show that multicomponent mixtures with chemical reactions are especially prone to spinodal instabilities, making vacuole formation more likely even when degradation occurs outside the droplet or production occurs in both phases.
Our work identifies spinodal instability as a generic mechanism for vacuole formation across a wide range of systems and perturbations.
This understanding provides a basis for designing molecular systems that exploit vacuole formation as a functional phenotype.

\section{Theory for vacuole formation}
To account for the various experimental settings with different perturbations leading to vacuole formation, we discuss binary and ternary mixtures, in a classical response setup. 
Specifically, the perturbation $h(t)$ is switched on at $t=0$ and we ask whether vacuoles form for $t>0$.
Moreover, we study two classes of how the perturbations $h(t)$ manifest in the system:
\begin{itemize}
    \item[\textbf{(i)}] \textbf{Perturbing interactions:} The perturbation $h(t)$ changes the strength of molecular interaction $\chi(h)$; see Appendix~\ref{appendix:thermo_quench} for details. Experimentally, molecular interactions are often modulated by changes in temperature $T$~\cite{Erkamp2023, Fritsch2021,Riback2017}, osmotic conditions~\cite{Moreau2020, Qamar2018,Krainer2021}, or pH~\cite{modi2026transient, Munder2016,Riback2017}, among others.

In general, any thermodynamic variable that affects molecular interactions can be captured by class \textbf{(i)}, provided its spatial dynamics is equilibrated faster than  diffusion of the macromolecules that scaffold the droplet.
The timescale $\tau_\text{q}$ over which $\chi(h)$ changes due to the time-dependent perturbation $h(t)$ is a crucial parameter of  class \textbf{(i)} and will affect vacuole formation. 

\item[\textbf{(ii)}] \textbf{Perturbing composition:} 
The perturbation $h(t)$ switches on a chemical reaction at time $t=0$ by modulating its chemical reaction rate coefficient  ${K}_i(h)=k_i \, h(t)$, in turn affecting the system composition. 
We investigate the case where $h(t)=\Theta(t)$ is a Heaviside function, yielding a continuous chemical turnover for $t>0$. The reaction is phase dependent, either occurring in the dilute phase ($k_i=k_{c,i}^\RN{1}$) or the dense phase ($k_i=k_{c,i}^\RN{2}$); see Appendix section~\ref{sect:appendix_rates} for details. 
Experimental examples of  class \textbf{(ii)} include condensates in which chemical reactions act directly on scaffold components, such as transcription or cleavage in synthetic DNA/RNA droplets~\cite{Saleh2023,verstraeten2025genetic}, fuel-driven cycles~\cite{Bergmann2023,Donau2020}, or enzymatic modification or degradation of scaffold proteins in biomolecular condensates~\cite{Monahan2017,Rai2018,Tsang2019}.

\end{itemize}
For each class, the system is initialized at phase equilibrium with a $\tanh$-like shaped interfacial profile $\phi_i(\mathbf{x},t)$ for each non-solvent component $i$ and a droplet of initial radius $R(0)$. 
For simplicity, we consider a two-dimensional system and quantify vacuole area $A_\text{vac}(t)$. 

\begin{figure*}[tb]
    \centering
    \makebox[\textwidth]{\includegraphics[width=\textwidth]{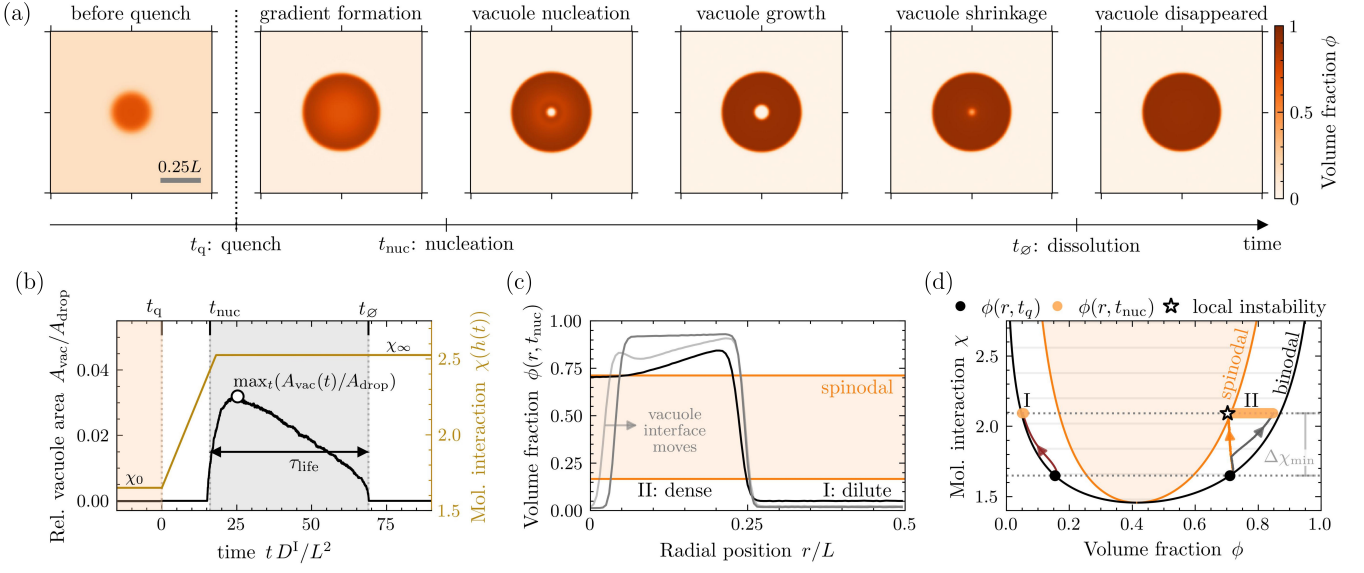}}
    \caption{\textbf{Physical mechanism of vacuole formation in a liquid droplet when changing interactions (class \textbf{(i)}):} 
    (a) Life-cycle of a droplet that is subjected to changes in interactions  $\chi(h(t))$ from $\chi_0$  to $\chi_\infty$ with a quench rate $k$; see golden curve in (b). 
    (b) When $\chi$ increases, the droplet (phase II) grows and develops an internal composition gradient (c). 
    This gradient triggers a local diffusive instability within the droplet interior when the droplet center touches the spinodal (see (d) for concentration trajectories of dilute phase 
    (red), droplet boundary (grey) and center position (orange) prior to instability), leading to the nucleation of a vacuole at $t=t_\text{nuc}$, which has a composition similar to the outside dilute phase I. 
    After nucleation, the vacuole initially grows, while it shrinks again on longer times (b), and eventually disappears, reaching a new thermodynamic equilibrium  related to the final interaction strength $\chi_\infty$.}
\label{fig_binary_temperature_quench}
\end{figure*}   

The spatial-temporal response of the volume fraction field $\phi_i(\mathbf{x},t)$  due to the perturbation $h(t)$ is governed for $t>0$ by stochastic irreversible 
thermodynamics~\cite{weberPhysicsActiveEmulsions2019d, julicher2024droplet, Zwicker2025DropletRegulation}:
\begin{align}
    \partial_t\phi_i &= \nabla\cdot\left(\frac{D_i\nu_i \phi_i\phi_S}{k_\text{B}T}\nabla\,\frac{\delta F(\chi(h))}{\delta \phi_i} + \bm{\xi}_i(\mathbf{x},\,t)\right)\nonumber\\[0.25em]
    &+ r_i(K_i(h)) + \eta_i(\mathbf{x},\,t)\,, 
    \label{eq:phidot}
\end{align}
where, neglecting cross-mobilities, $D_i=D_i(\phi_S)$ is a composition-dependent diffusion coefficient of component $i$ with constant molecular volume $\nu_i$, $\phi_S$ is the solvent volume fraction, and $k_\text{B}T$ is the thermal energy. 
The stochastic vector field $\bm{\xi}_i$ describes fluctuations in the diffusive flux, whereas the stochastic scalar field $\eta_i$ describes fluctuations associated with the chemical reactions.
Both are taken to be uncorrelated white-noise fields with zero mean, obeying the fluctuation-dissipation theorem.
Diffusive fluxes evolve according to spatial gradients in the chemical potentials $\mu_i=\nu_i\delta F/\delta \phi_i$, where $F(\chi)$ is the system free energy  that depends on the molecular interaction strength $\chi(h)$ that gets perturbed in class \textbf{(i)} (see Appendix~\ref{appendix:free_energy} for details). 
In class \textbf{(ii)}, chemical pathways are accounted for by the reaction flux $r_i(K_i(h))$ that follows a simple mass-action law kinetic, with $r_i(K_i(h))$ proportional to the volume fraction of the reacting component(s) (details see Appendix~\ref{sect:appendix_rates}). 
The discretized and rescaled equations can be found in Appendix~\ref{appendix:rescaled_numeric}.
Parameters are chosen consistent with experimentally observed droplets and vacuoles with sizes larger than the microscopic interfacial width, while  still being small enough such that fluctuations matter (see Table~\ref{tab:simulation_parameters} for the chosen length scales and parameters).

\section{Results} In the following, we show that in both classes \textbf{(i, ii)}, vacuoles form when spatial gradients in the droplet composition drive part of the droplet into the spinodal region of the thermodynamic phase diagram. 
Entering the spinodal region has a kinetic requirement: in class \textbf{(i)}, the spinodal boundary shifts with $\chi$ faster than diffusion can homogenize the droplet, whereas in class \textbf{(ii)}, reaction fluxes drive the local composition into the spinodal faster than diffusive fluxes can relax it.

\subsection*{Class (i): Vacuole formation by interaction change} 
For $t>t_\text{q}$, the droplet is subject to an increase in interaction strength $\chi(h(t))$ from $\chi(h(0))\equiv\chi_0$ to $\chi(h(\infty))\equiv\chi_\infty$ on a time scale $\tau_\text{q}=k^{-1}$, a process commonly referred to as quenching, where $k$ is the quench rate.
For an illustration of $\chi(h)$, see Fig.~\ref{fig_binary_temperature_quench}(b) (golden curve), with more details of the quenching protocol in Appendix~\ref{appendix:thermo_quench}. 
For sufficiently small diffusivity in the dense phase ($D^\RN{2}\ll kR ^2$), a pronounced concentration gradient emerges during quenching (Fig.~\ref{fig_binary_temperature_quench}(a) from left to right). 
When this concentration gradient is pronounced enough, the composition at the center of the droplet becomes locally unstable, resulting in vacuole nucleation at $t=t_\text{nuc}$. 
The vacuole grows quickly in size, reaching its maximum size slightly after the changing interactions $\chi$ saturate; see Fig.~\ref{fig_binary_temperature_quench}(b) showing the vacuole area.
For longer times, the vacuole slowly shrinks until it eventually disappears at time $t_\varnothing$, reaching the system's new equilibrium state corresponding to the saturation value of interaction strength $\chi_\infty$.
The time from vacuole formation to its dissolution is defined as the vacuole life-time $\tau_\text{life}=t_\varnothing - t_\text{nuc}$; see black arrow in Fig.~\ref{fig_binary_temperature_quench}(b).

\begin{figure*}[tb]
    \centering
    \makebox[\textwidth]{\includegraphics[width=\textwidth]{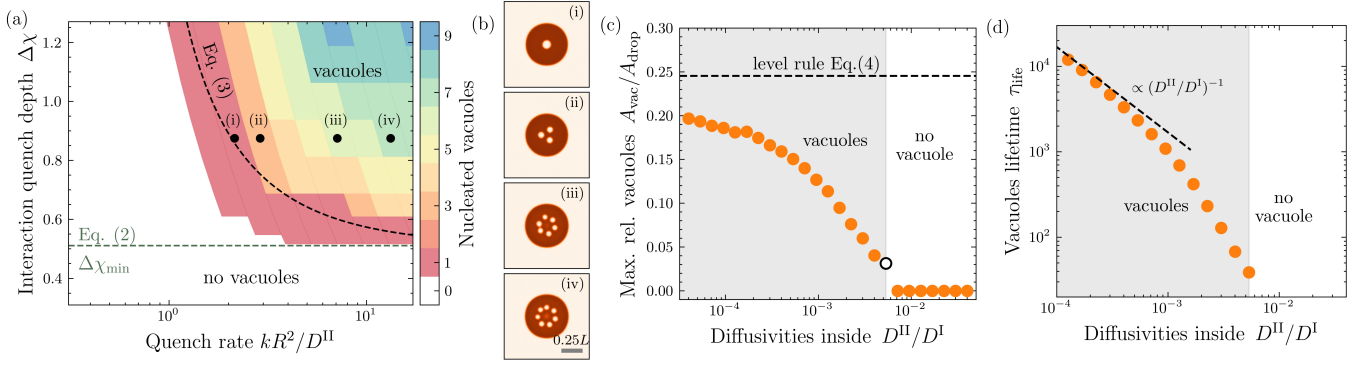}}
    \caption{\textbf{Formation of many vacuoles in a liquid droplet when changing interactions (class \textbf{(i)}):}
    (a) Vacuole formation requires both sufficient quench depths $\Delta \chi > \Delta\chi_\text{min}$ (Eq.~\eqref{eq:min_chi}) and large enough quench rates $k > k_\text{crit}$ (Eq.~\eqref{eq:threshold}).
    This competition is confirmed by our analytics (Eq.~\eqref{eq:max_vac_area}, dashed line in (a)) that agrees well with the numerical results for a single vacuole. 
    (b) Snapshots of droplets (i)--(iv) at fixed quench depth $\Delta \chi$ but increasing quench rate $k R^2/D^\text{II}$ show the transition from one to seven vacuoles.
    We note that a single vacuole always forms close to the center since the center undergoes the instability first, with the gradient being monotonous from the center to the interface.  
    (c) Maximum relative vacuole area $A_\text{vac}/A_\text{drop}$ as a function of relative diffusivity, $D^\text{II}/D^\text{I}$ between inside (II) and outside (I) of the droplet. 
    The maximal total vacuole area occurring at low $D^\text{II}/D^\text{I}$ is well captured by a lever-rule for the vacuole phase, Eq.~\eqref{eq:max_Vvac}. 
    Open white symbol corresponds to the parameters used in Fig.~\ref{fig_binary_temperature_quench}. 
    (d) Vacuole lifetime $\tau_\text{life}$ as a function of the relative diffusivity.
    The scaling of the lifetime with diffusivity inside,  $\tau_\text{life} \propto 1/D^\RN{2}$, explains  long-lived vacuoles in experiments, noting that a low diffusivity inside is also a requirement for their formation.}
    \label{fig:2}
\end{figure*}

Vacuoles typically nucleate right after the radially symmetric concentration profile $\phi(r,t=t_\text{nuc})$ intersects with the (dense) spinodal line, which is shown by the orange line in Fig.~\ref{fig_binary_temperature_quench}(c). 
For concentrations above the spinodal, with concentration fluctuations ($\eta_i$ in Eq.~\eqref{eq:phidot}), we could not observe the formation of a mesoscopic vacuole in our simulations (see Appendix~\ref{sec:app:thermal_fluctuation}, Fig.~\ref{fig_app:Stocastic_runs}).
This observation suggests that vacuole formation is dominantly set by the deterministic contribution in Eq.~\eqref{eq:phidot}, while fluctuations affect only where inside the droplet and slightly when vacuoles get nucleated. 
These deterministic contributions contain the local spinodal instability that can be understood through the phase diagram before and after changing the interactions $\chi(h)$.
According to the phase diagram (Fig.~\ref{fig_binary_temperature_quench}(d)), increasing the molecular interactions $\chi(t)$ enlarges the spinodal domain of  unstable concentrations  (shown in orange).
When the quenching rate $k$ is fast compared to diffusion inside the droplet, the droplet's composition cannot equilibrate to the new equilibrium at $\chi(t)$.
The concentrations within the droplet lag behind such that they may intersect with the dense spinodal (see black star and orange arrows in Fig.~\ref{fig_binary_temperature_quench}(d)). 
This lag is larger when $k\gg D^\RN{2}/R^2$, where $R$ is the droplet radius and $D^\RN{2}$ is the diffusion coefficient inside the droplet phase, thereby facilitating vacuole formation.

In almost all of our studies, vacuoles form inside the droplet, while the dilute phase remains stable, with no new droplets forming. This asymmetry reflects two typical experimental conditions of biomolecular condensates \textit{in vitro}:
First, such condensates are composed of macromolecules as scaffolds (polymers, proteins, RNAs, etc.), implying a much lower diffusion coefficient inside compared to outside ($D^\RN{1}\gg D^\RN{2}$). 
These asymmetric diffusivities make the dilute phase evolve along the dilute binodal, maintaining its stability locally, while the dense phase can deviate if the interaction changes fast enough (see trajectories in Fig.~\ref{fig_binary_temperature_quench}(d)).
Second,  the size of macromolecular scaffolds $N$ exceeds the solvent ($N\gg1$), which is typically an aqueous buffer composed of small molecules (water, salt ions). 
This asymmetry in molecule sizes makes the dilute branch of the binodal vary less with the interaction parameter $\chi$ than the dense branch. 
Thus, changes in interactions cause the concentration inside the droplet to change much more than outside, thereby favoring vacuole formation inside the droplet over droplet nucleation outside.

To nucleate vacuoles, the quench in interaction strength $\chi(t)$ must be both fast and strong.
If the quench depth $\Delta \chi$ is too shallow, a droplet can never form vacuoles as the initial droplet composition is stable even after the quench.
The minimal quench depth $\Delta \chi_\text{min}$  for vacuole formation can be determined from the thermodynamic phase diagram (derivation see Appendix~\ref{appendix:analytics}),
\begin{equation}
\label{eq:min_chi}
    \Delta\chi_{\min} = \frac{(N-1)\phi_0^\RN{2}+1}{2N\phi_0^\RN{2}(1-\phi_0^\RN{2})} - \chi_0\,,
\end{equation}
where $\phi_0^\RN{2}$ denotes the droplet composition before the quench.
This prediction agrees well with the numerical data displayed in Fig.~\ref{fig:2}(a), see green dashed line.
For vacuole formation, additionally, the interaction quench must be fast compared to the time-scale of diffusive relaxation ($D^\RN{2} \ll kR^{2}$) for the concentration at the droplet center to lag behind the concentration at the droplet boundary. Vacuoles form when $k>k_\text{crit}$, with the critical quench rate 
\begin{equation}
\label{eq:threshold}
    k_\text{crit} = \frac{D^\RN{2}}{2R^2}\left(\frac{\phi^\RN{2}_\infty-\phi^\text{II}_0}{\phi^\RN{2}_\text{spin}(\chi_\infty)-\phi^\RN{2}_0}\right) \, . 
\end{equation}
Here, $\phi_\text{spin}^\RN{2}$ is the dense spinodal composition and $\phi_\infty^\RN{2}$ is the dense-phase equilibrium composition after the quench at $\chi_\infty$.
In Eq.~\eqref{eq:threshold}, the $\phi$-fraction quantifies the proportion of the meta-stable region compared to the total compositional change of the droplet. 
In Appendix~\ref{appendix:analytics}, the expression is derived by approximating the binodal and spinodal compositions, finding a closed-form expression that agrees well with the numerical results shown by a black dashed line in Fig.~\ref{fig:2}(a).

For faster and stronger interaction quenches ($k\geq k_\text{crit}$ and $\Delta \chi > \Delta\chi_\text{min}$), more vacuoles form inside the droplet; a trend that was also experimentally observed  using temperature quenches~\cite{Erkamp2023}.
In our numerical studies, the maximal number of observed vacuoles transitions from one (i) to, e.g., seven (iv) for a larger quenching rate $k$, as shown in Fig.~\ref{fig:2}(b). 
For a given quench depth $\Delta\chi$, the number of vacuoles saturates for large quench rates $kR^2/D^\RN{2}$.

Concomitant with the vacuole number, the total maximal vacuole area $A_\text{vac}$ increases and saturates as the dense phase diffusivity $D^\RN{2}/(kR^2)$ becomes small (Fig.~\ref{fig:2}(c)). 
This result highlights that vacuole formation requires a pronounced  asymmetry between the slower transport inside relative to outside.
Since the total droplet material is conserved as it is redistributed between the dense phase and the dilute vacuole, mass balance fixes their relative areas, yielding a lever rule for the maximum possible relative vacuole area; (derivation in Appendix~\ref{appendix:analytics}),
\begin{equation}   
\label{eq:max_vac_area}
\text{max}_t\left(\frac{A_\text{vac}(t)}{A_\text{drop}}\right) = \dfrac{\frac{1}{1+\sqrt{N}}+\sqrt{\frac{3(\chi_0-\chi_*)}{2\chi_*^2\sqrt{N}}}-1+e^{-\gamma_N-\chi_\infty}}{e^{-1-N(\chi_\infty-1)}+e^{-\chi_\infty-\gamma_N}-1} \, , 
\end{equation}
where we have defined $\gamma_N = 1-{1}/{N}$. 
We find that vacuoles approach the droplet size,
$A_\text{drop}- \max_t\left(A_\text{vac}(t)\right)\propto N^{-1/4}$,
for long polymers phase-separating from smaller buffer molecules ($N\gg1$). 
The analytic lever rule (Eq.~\eqref{eq:max_vac_area}) agrees well with the saturation level obtained from numerics (Fig.~\ref{fig:2}(c)).

\begin{figure*}[tb]
    \centering
    \makebox[\textwidth]{\includegraphics[width=\textwidth]{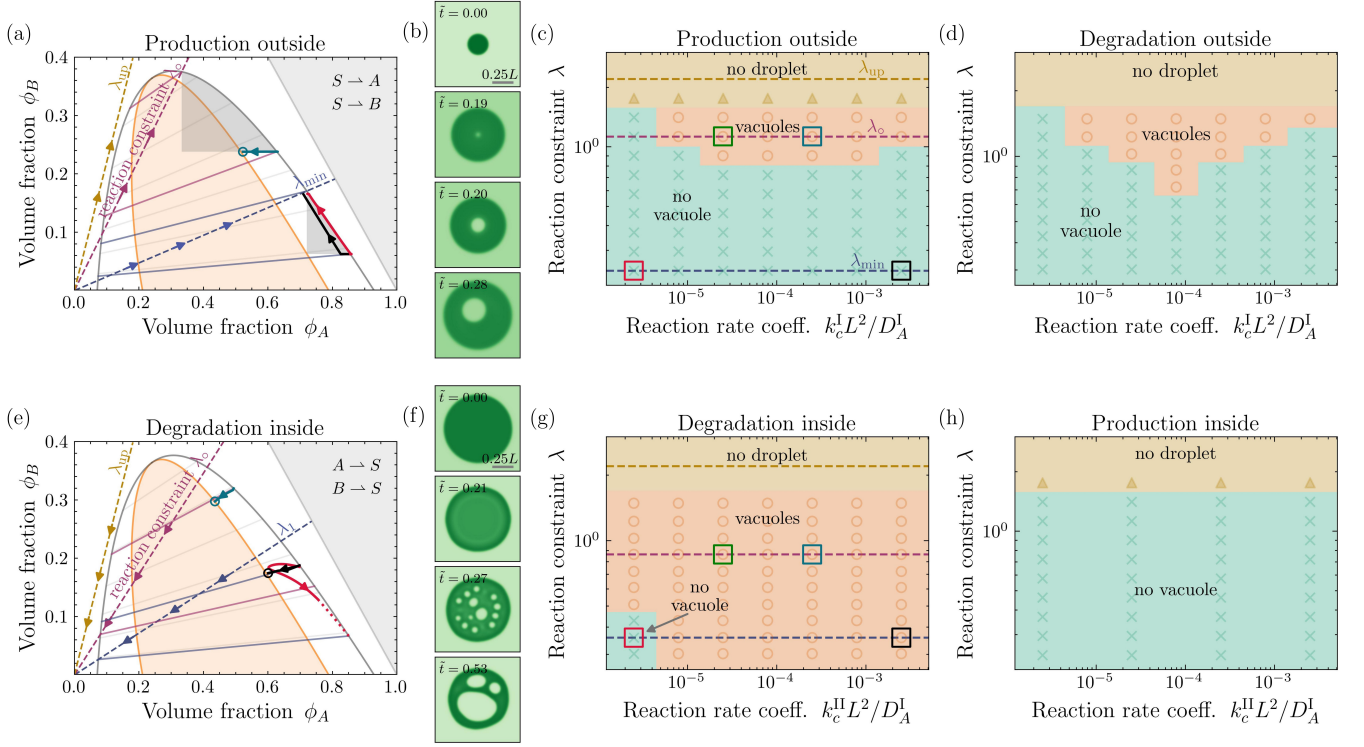}}
    \caption{\textbf{Chemical reaction schemes can trigger vacuole nucleation in ternary systems:}
    We characterize the production and degradation reactions (see Eq. \eqref{eq:ternary_reaction}) in two different scenarios, outside (a-d) or inside (e-h) the droplet. 
    Phase-space trajectories are displayed in (a) and (e).
    The evolution of a droplet during production outside and degradation inside is shown in (b) and (f), respectively, for the cases highlighted by the green boxes in (c) and (g).
    We vary the reaction constraint $\lambda$ and the reaction-diffusion length scale for all four cases, and summarize the outcome (c,d,g,h), where vacuole formation is indicated by orange circles and the absence of vacuoles by green crosses. 
    Vacuoles are most readily observed when degradation occurs within the droplet.}
    \label{fig3:reactions_kin}
\end{figure*}

After vacuoles are nucleated, they grow in time before shrinkage occurs on longer times (see Fig.~\ref{fig_binary_temperature_quench}(a, frame 3-4) for a single vacuole).
This vacuole growth after nucleation poses the question of what drives the increase of an unfavorable larger vacuole interfacial area?
When a single vacuole is nucleated close to the droplet center,  a parabolically-shaped  concentration profile spans from the vacuole interface to the outer droplet interface (light gray in Fig.~\ref{fig_binary_temperature_quench}(c)).
When a vacuole grows, this internal concentration gradient between the inner and outer interface homogenizes. 
It is this associated decrease in  free energy that drives vacuole growth after nucleation. 
The reason for the growth of the vacuole is therefore equivalent to the case of a droplet in a supersaturated external phase, just with the compositions switched, where the vacuole containing the dilute phase grows inside the dense phase. 
Once this internal concentration gradient is homogenized (Fig.~\ref{fig_binary_temperature_quench}(a, frame 4)), the vacuole   shrinks and eventually dissolves.
These kinetics result from the classical driver of coarsening: lowering the free energy associated with the vacuole interface inside the roughly homogeneous dense phase.

The life-time $\tau_\text{life}$ of a vacuole -- from nucleation to its dissolution --  increases with lower diffusivity in the dense phase $D^\RN{2}$ (Fig.~\ref{fig:2} (d)). 
For sufficiently slow diffusivity inside the droplet, the lifetime is rate-limited by scaffold molecules  diffusing inside the droplet, 
$\tau_\text{life} \propto 1/{D^\RN{2}}$ (dashed line). 
This behavior indicates that droplets with many vacuoles (i.e., large total vacuole area $A_\text{vac}$) also yield a longer lifetime $\tau_\text{life}$.
These long lifetimes may explain the stability of vacuoles over typical laboratory time scales of hours~\cite{li2025static,Bergmann2023} to days~\cite{banerjee2017reentrant,verstraeten2025genetic}.

\subsection*{Class (ii): Vacuoles formation by chemical reactions}
\label{sect:chemical_reactions}
Here we discuss how vacuoles form through chemical reactions described by the reaction flux $r_i\propto K_i(h(t))$ that is suddenly switched on at $t=0$ in Eq.~\eqref{eq:phidot}, while keeping all the interactions $\chi$ fixed. 
Note that the system is not at a steady chemical state after switching on the perturbation. 
In a simple binary system, the spinodal domain is a one-dimensional range of concentration values. 
Thus, the only reaction scheme that can trigger vacuole formation is degradation of droplet material within the droplet (phase II) with $r^\text{II}=-k_c \phi^\RN{2}$.
Only for degradation is there a decaying profile from the interface toward the droplet center, allowing it to trigger the spinodal instability locally.
For a systematic study of the four possible reaction pathways (production/degradation, inside/outside) in binary systems, see Appendix~\ref{sec:app:binary_react}.

To draw conclusions about the propensity for vacuole formation in actual chemical systems, we considered multi-component systems subject to conservation laws of mass. 
A minimal yet not oversimplified system is a ternary mixture, in which one molecular component is converted to another through unidirectional chemical transitions, while conserving mass. 
For this ternary systems,  various phase diagrams are possible, leading to richer scenarios of vacuole formation for different chemical processes than for the oversimplified binary systems.
In the following, we investigate a ternary system subject either to unidirectional production reactions ($r_+$), or to unidirectional degradation reactions ($r_-$): 
\begin{subequations}
\label{eq:ternary_reaction}
\begin{align}
    r_+:\qquad
    S &\rightharpoonup A,\quad \text{and}
    &
    S &\rightharpoonup B,
    \label{eq:production_reactions}
    \\
    r_-:\qquad
    A &\rightharpoonup S,
    \quad \text{and}
    &
    B &\rightharpoonup S.
    \label{eq:degradation_reactions}
\end{align}
\end{subequations}
For $t>0$, chemical reactions are switched on with the reaction rate proportional to the volume fraction of the reacting components:
\begin{align}
    \text{production:}\qquad r_A &= k_{A}\phi_S\,,\quad \quad 
    r_B= k_{B}\phi_S\,, \\
    \text{degradation:}\qquad r_A &=  -k_A\phi_A\,,\quad \, r_B= -k_B\phi_B\,,
\end{align}
following simple mass-action-law-like kinetics known for dilute systems. 
For the production reactions (Eq.~\eqref{eq:production_reactions}), the two reactions occur with different, phase-dependent rate coefficients $k_i(\phi_S)$ ($i=A,B$) whose ratio is fixed, $k_B/k_A = \lambda$, whereas the degradation reactions (Eq.~\eqref{eq:degradation_reactions}) share a single rate coefficient ($k_A = k_B$). The system is initialized with $\phi_B/\phi_A = \lambda$. We therefore refer to $\lambda$ as a reaction constraint, since it sets the average ratio of $B$ to $A$ produced or degraded in the system.
For each value of $\lambda$, the initial droplet is chosen from the binodal composition along the corresponding reaction-constraint line $\lambda$. 
Specifically, we select the coexistence state for which the droplet occupies $5\%$ of the total system area for production reactions and $50\%$ for degradation reactions.

A key finding is that chemical reaction-driven vacuole formation
is a rather generic phenomenon in multi-component mixtures.
This is evident in the possibility of vacuole formation across most scenarios of production/degradation occurring inside/outside droplets. 
We characterize and explore the propensity of reaction-driven vacuole formation by varying the reaction constraints $\lambda$ and reaction rate coefficients $k_c$ (Fig.~\ref{fig3:reactions_kin}(c,d,g,h)).
We find vacuole formation can also occur in scenarios beyond degradation inside; 
contrary to binary systems (Appendix \ref{sec:app:binary_react}, Fig.~\ref{fig_binary_chemical_reaction}).
The exception is production (+) occurring in the dense phase, where vacuoles cannot form because the deviation from the dense binodal is away from the spinodal. 
In contrast, for degradation inside,  the trajectory of the center dense phase concentrations can intersect with the spinodal line, leading to vacuole formation (Fig.~\ref{fig3:reactions_kin}(e,f,g)).

\begin{figure*}[tb]
    \centering
    \includegraphics[width=0.825\linewidth]{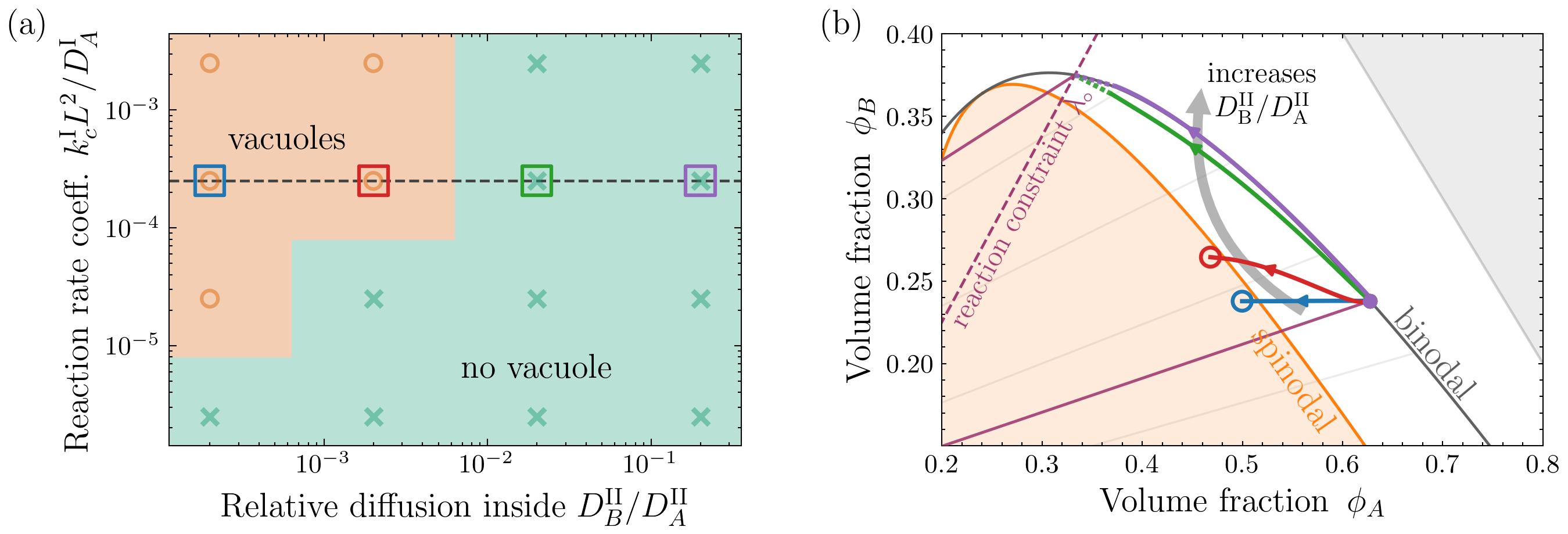}
    \caption{\textbf{Asymmetric diffusion drives spinodal instability and vacuole nucleation in multi-component droplets.}
    (a) Vacuole formation (orange circles; no vacuoles are indicated in green crosses) does not necessarily occur when the reaction-diffusion length scale $\sqrt{D_i^\RN{2}/k_c}$ is small relative to the droplet size $R$.
    (b)  To intersect with the spinodal and trigger the vacuole instability, the  relative diffusivities inside between the components, $D_B^\RN{2}/D_A^\RN{2}$, need to be different, which depends on the phase diagram and the spinodal shape. 
    For the diagram shown in (b), $D_B^\RN{2}/D_A^\RN{2}\ll 1$ to ensure a rather horizontal trajectory (see blue) of the composition at the droplet center.}
    \label{fig:4asym_diffusion}
\end{figure*}

Vacuoles are more prevalent in multi-component systems since the concentration profile has additional degrees of freedom by which the system can deviate from phase equilibrium.  
For reactions occurring in the dilute phase, the dense-phase composition evolves through diffusive exchange with the dilute phase. 
As the reactions drive the system's average composition along the reaction-constraint line $\lambda$, the dense-phase composition is drawn toward the attractor set by the intersection of this line with the dense branch of the binodal (for production $+$) or the tie-line intersecting it on the dilute branch (for degradation $-$). 
When $A$ and $B$ diffuse at different rates, they relax on different timescales: in the extreme $D_A^\RN{2}\gg D_B^\RN{2}$, the faster component adjusts first while the slower one remains essentially fixed. 
Spanning the accessible diffusivity ratios, the composition trajectories are therefore confined to a triangular domain (grey region in Fig.~\ref{fig3:reactions_kin}(a)), bounded by the constraint line $\lambda$ and the initial and final dense-phase compositions. 
A thermodynamic prerequisite for vacuole nucleation is that this accessible domain overlaps with the spinodal: only then can the droplet-center composition deviate far enough from the binodal to cross into the spinodal region and trigger the instability, as illustrated by the green trajectory for reaction constraint $\lambda_\circ$.

To fully explore vacuole formation in reacting multi-component systems, we studied the four different reaction cases (Eqs.~\eqref{eq:ternary_reaction}) for different types of phase diagrams (Appendix~\ref{sec:app:vary_bino}, Fig.~\ref{fig:extra_state_diagram}). 
Overall, when degrading in the dense phase, irrespective of the specific phase diagram, the system is particularly prone to vacuole formation when varying the reaction constraint $\lambda$ and the reaction rate coefficient $k_c^\text{II}$. 
In contrast, other reactions can only give rise to tiny triangular domains in the respective phase diagrams because the reaction constraint is roughly parallel to the tie lines    
(Appendix Fig.~\ref{fig:state_diagram_with_ic}).

The relative diffusion between the reaction components $A$ and $B$ determines the initial directionality of the droplet composition in the ternary phase diagram. 
Therefore, to nucleate vacuoles in multi-component systems requires the relative diffusion coefficient $D_A^\RN{2}/D_B^\RN{2}$ to be such that the composition moves towards the spinodal.
For the phase diagram depicted in Fig.~\ref{fig3:reactions_kin}(a), $\phi_A^\RN{2}$ should decrease while $\phi_B^\RN{2}$ remains roughly constant to move towards the spinodal, which occurs if $D_B^\text{II}\ll D_A^\text{II}$.
Thus, the reaction-diffusion length scale $\sqrt{D_i^\RN{2}/k_c}$ should be small compared to the droplet, but if $D_A^\RN{2}\simeq D_B^\RN{2}$, the composition will not necessarily deviate from the binodal (green and purple trajectories in Fig.\ref{fig:4asym_diffusion}(b)). 
Thus, the effects of asymmetric diffusion are evident in Fig.~\ref{fig:4asym_diffusion}(a,b), showing that a small reaction-diffusion length scale  on its own is not sufficient for vacuole formation, but must be accompanied by an asymmetry in the diffusion coefficients for vacuoles to form.

\section{Conclusion} 

The widespread observation of vacuoles~\cite{SchmidtRohatgi2016,Boeynaems2019,Alshareedah2020,Li2025, modi2026transient,Moreau2020,Yewdall2021,modi2026transient, Saleh2023,verstraeten2025genetic,
Erkamp2023, Bartolucci2021, giessler2026growth, Bergmann2023, bauermann2023formation} in chemically very different experimental systems exposed to very different processes that perturb the droplet composition in space and time suggest a common mechanism. 
Indeed, our theory shows that large droplets with particularly low diffusivity inside can easily form vacuoles through a local spinodal instability.
We elucidate why macromolecules forming liquid droplet phase-separated from smaller solvent molecules are particularly prone to vacuole formation -- a common setting for biomolecular condensates~\cite{Banani2017,Lyon2021,Fare2021} and coacervates~\cite{Abbas2021,Rumyantsev2021,Donau2020,Moreau2020} \textit{in vitro}, where vacuoles have been observed frequently with remarkably large life times of hours to many days. 
Our theory also explains that the long-lived nature of vacuoles is related to their emergence: only when diffusion is slow enough inside, vacuoles form, which also gives rise to the large vacuole lifetime. 
Specifically, experimentally evidenced diffusivity in the order of $10^{-2} \mu m^2/s$  inside droplet (and even lower) with a typical droplet size of, say $20 \, \mu m$, gives life times of many hours up to a day.

Revealing the generic mechanism for vacuole formation in liquid droplet paves the way for efficient strategies of molecular design, exploiting the vacuole phenotype for functional engineering.
One key phenotypic property of many vacuoles is the increase in total surface area between the dilute and dense droplet phase. 
For example, for ten, $1\, \mu m$-thick vacuoles along the droplet diameter of $20 \, \mu m$, the relative excess area in three dimensions is around a factor of three, suggesting significant amplification of surface-mediated processes. 
While coarsening generically decreases the excess area when approaching thermodynamic equilibrium, vacuole formation offers a route to suddenly create excess interface area, even in passive systems, affecting  physicochemical properties of the system upon vacuole formation. 
A striking example is interface-mediated catalysis~\cite{Straathof2003,Stroberg2018,Smokers2024} for which the catalytic activity enhances with the excess vacuole interfacial area.

We speculate that many vacuoles inside a droplet make the droplet prone to fission, for example in the presence of shear flow or starving fuel-driven chemical systems~\cite{wenisch2025toward}. 
Since vacuoles can form so easily in condensates composed of macromolecules, they provide a simple yet robust mechanism for protocells to produce daughter droplets for \textit{de novo} life~\cite{Zwicker2017,Yewdall2018,haugerud2026theory}.
Following early ideas from Oparin and Haldane~\cite{Oparin1938,Haldane1929}, the generic character of the vacuole instability also suggests relevance for the molecular origin of life.

\begin{acknowledgments}
We thank A.\ Thatte, H.D.\ Vuijk, 
E.\ Araspin,  and L.\ Silva-Dias for the feedback on the manuscript and fruitful discussions on the subject.  
C.\ Weber and P.\ Jaiswal thank the University of Augsburg for financial support via the research program ``Forschungspotentiale besser nutzen!''. 
C.\ Weber (Project number 521256690) and K.\ G\"opfrich (Project number 537091438) thank the TRR 392: Molecular evolution in prebiotic environments for support. 
The authors gratefully acknowledge the resources on the LiCCA/ALCC HPC cluster of the University of Augsburg, co-funded by the Deutsche Forschungsgemeinschaft (DFG, German Research Foundation) – Project-ID 499211671.
K.\ G\"opfrich acknowledges the Cluster of Excellence SynthImmune, the Human Frontiers Science Program (RGPO03I2023), the Alfried Krupp F\"orderpreis and the ERC Starting Grant ``ENSYNC'' (No. 101076997).
C.\ Weber acknowledges the European Research Council (ERC) under the European Union’s Horizon 2020 research and innovation program (Fuelled Life, Grant No. 949021).
\end{acknowledgments}

\appendix

\section{Perturbation setup affecting molecular interactions and composition}
\label{appendix:thermo_quench}

As the diffusion of temperature is typically orders of magnitude faster than the diffusion of macromolecules~\cite{julicher2024droplet}, we treat $\chi(h(t))$ as a homogeneous control parameter.
The perturbation is applied at $t=0$ and changes the interaction parameter from its initial value $\chi(h(0))\equiv\chi_0$ to a final plateau value $\chi(h(\infty))\equiv\chi_\infty$. 
We write this dependence as
\begin{equation}
    \label{eq:chi_perturbation}
    \chi(h(t))
    =    \left(\chi_\infty-\chi_0\right)h(t)
    +
    \chi_0\,,
\end{equation}
where $h(t)$ specifies the temporal protocol of the perturbation.
For a linear quench of duration $\tau_\text{q}$, we choose
\begin{equation}
    \label{eq:chi_quench_protocol}
    h(t)
    =
    \begin{cases}
    0\,,
    & t<0\,, \\[0.4em]
    \dfrac{t}{\tau_\text{q}}\,,
    & 0\leq t<\tau_\text{q}\,, \\[0.4em]
    1\,,
    & t\geq\tau_\text{q}\,.
    \end{cases}
\end{equation}
This quench protocol is illustrated in Fig.~\ref{fig_binary_temperature_quench}(b) of the main text.

For the reaction-driven perturbation, the chemical process is activated through the perturbation-dependent reaction-rate coefficient
\begin{equation}
\label{eq:reaction_rate_switch_protocol}
    K_i(h)=k_i(\phi_S)\,h(t)\,,
\end{equation}
where $k_i(\phi_S)$ describes the local composition dependence of the reaction rate, while $h(t)$ controls its temporal activation.
Since chemical reactions already come with an intrinsic time-scale ($k_i(\phi_S)^{-1}$), we consider an instantaneous switch at $t=0$:
\begin{equation}
    h(t)=\Theta(t)\,.
\end{equation}
Accordingly, the chemical process is inactive for $t<0$ and proceeds continuously for $t\geq 0$. The specific composition dependence of $k_i(\phi)$ is defined in Eq.~\eqref{eq:k_c_switch}.

\section{Liquid mixture free energy}
\label{appendix:free_energy}

We describe the liquid mixture using the volume fractions fields of the components
composition fields $\phi_i(\mathbf r,t)$. 
The thermodynamics of the mixture is encoded in the coarse-grained free-energy functional
\begin{equation}
\label{eq:free_energy_total}
    F[\{\phi_i\},\{\nabla \phi_i\}]
    =
    \int \mathrm{d}V
    \left[
        f(\{\phi_i\},\chi_{ij})
        +
        \sum_{i=0}^{M} \frac{\kappa_i}{2\nu_0}|\nabla\phi_i|^2
    \right]\,.
\end{equation}
Here,  the square-gradient term penalizes spatial variations in composition and therefore gives a finite interfacial cost between coexisting phases.
We use the Flory--Huggins free-energy density~\cite{weberPhysicsActiveEmulsions2019d}
\begin{equation}
\label{eq:free_energy_density}
\begin{aligned}
    f(\{\phi_i\},\chi_{ij})
    &=
    \frac{k_\mathrm{B}T}{\nu_0}
    \Bigg[
        \sum_{i=0}^{M} \frac{\phi_i}{N_i}\log\phi_i
        +
        \sum_{i=0}^{M} \omega_i\,\phi_i
        \\
        & \quad +
        \frac{1}{2}\sum_{i,j=0}^{M}
        \chi_{ij}\,\phi_i\phi_j
    \Bigg]\,.
\end{aligned}
\end{equation}
Here, $N_i = \nu_i/\nu_0$ denotes the molecular volume of component $i$ relative to the reference volume $\nu_0$. 
The logarithmic term represents the entropy of mixing, the parameters $\omega_i$ correspond to internal energies, and the interaction parameters $\chi_{ij}$ quantify pairwise interactions between species.

The chemical potential associated with the conserved composition field is obtained from the functional derivative of the free energy. 
For an incompressible mixture, where the solvent volume fraction is given by $\phi_S = 1-\sum_{k=1}^{M}\phi_k$, we define
\begin{equation}
\label{eq:chem_pot_definition}
    {\mu}_{i}
    =
    \nu_i
    \left(
        \frac{\delta F}{\delta \phi_i}
    \right)_{T,V,\phi_{j\neq i}}
    \, .
\end{equation}
Using Eq.~\eqref{eq:free_energy_total}, this gives
\begin{align}
    \mu_i
    &=
    k_\mathrm{B}T
    \left(1+\log\phi_i\right)
    -
    k_\mathrm{B}T N_i
    \left[
        1+
        \log
        \left(
            1-\sum_{k=1}^{M}\phi_k
        \right)
    \right]
    \nonumber\\
    &\quad
    +
    k_\mathrm{B}T N_i
    \left(\omega_i-\omega_0\right)
    +
    k_\mathrm{B}T N_i\chi_{i0}
    \left(
        1-\sum_{k=1}^{M}\phi_k
    \right)
    \nonumber\\
    &\quad
    +
    k_\mathrm{B}T N_i
    \sum_{k=1}^{M}
    \left(
        \chi_{ik}-\chi_{0k}
    \right)
    \phi_k
    -
    N_i\kappa_i\nabla^2\phi_i .
    \label{eq:exc_chem_pot_2}
\end{align}

\section{Rate scheme and rates for a ternary system}
\label{sect:appendix_rates}

In the following, we discuss the composition-dependent reaction rate after the chemical reaction is switched on ($t>0$); for protocol see Appendix~\ref{appendix:thermo_quench}.
The mixture is subjected to chemical reactions that can either produce droplet material through solvent as $S \rightharpoonup A$ and $S\rightharpoonup B$ with species specific reaction rates as
\begin{align}
    r_A = +\phi_S k_A(\phi_S),\quad
    r_B = +\phi_S k_B(\phi_S)\,,
\end{align}
or degrade droplet material to solvent $A \rightharpoonup S$ and \\$B \rightharpoonup S$, with
\begin{align}
    r_A = -\phi_A k_A(\phi_S),\quad
    r_B = -\phi_B k_B(\phi_S).
\end{align}
In both cases, the rates satisfy the constraint $r_B/r_A = \lambda$. 
The same reaction constraint can be achieved through material exchange with an external  reservoir \cite{jaiswal2026harvesting,haugerud2024nonequilibrium}.
The reactions are localized either in the solvent-rich outside phase ($\mathrm{I}$) or in the solvent-poor inside phase ($\mathrm{II}$) by using a smooth, composition-dependent reaction rate as
\begin{equation}
\label{eq:k_c_switch}
    k_i(\phi_S) =
    k_{c,i}\left[
    \bigg(\frac{1+\alpha}{2}\bigg)
    \pm
    \bigg(\frac{1-\alpha}{2}\bigg)
    \tanh\bigg\{
    \frac{\phi_S-\phi_0}{\delta_\phi}
    \bigg\}
    \right] \, .
\end{equation}
The `$+$' sign localizes the reaction predominantly in the solvent-rich phase, whereas the `$-$' sign localizes it in the solvent-poor phase.
Here, $k_{c,i}$ is the reaction rate coefficient in the phase where reactions are occurring. 
For the production reaction, we set $k_{c,A}=k_c$ and choose $k_{c,B}=\lambda k_c$, such that $k_{c,B}/k_{c,A}=\lambda$. 
For the degradation reaction, both components are degraded with the same rate coefficient, $k_{c,A}=k_{c,B}=k_c$.
In Eq.~\eqref{eq:k_c_switch}, $\alpha$ sets the ratio between the reaction-rate coefficient in the non-reactive phase and that in the reactive phase.
In all simulations shown here, we set $\alpha=0$, so that the reaction is fully localized to either inside (dense phase) or outside (dilute phase).
$\phi_0$ sets the switching threshold, and $\delta_\phi$ controls the width of the transition.

\section{Discretized and rescaled dynamical equations and numerical method}
\label{appendix:rescaled_numeric}

We rescale lengths by a reference length scale $L_0$ and time by the corresponding diffusive time scale
  $  \tau
    =
    {L_0^2}/{D_0} $,
\begin{equation}
    \label{eq:define_L_and_t}
    \widetilde{\mathbf r}
    =
    \frac{\mathbf r}{L_0} \, ,
    \qquad
    \tilde{\nabla }
    =
    \nabla {L_0} \, ,
    \qquad
    \widetilde t
    =
    \frac{t}{\tau}
    =
    \frac{tD_0}{L_0^2} \, ,
\end{equation}
where $D_0$ is a reference diffusion coefficient. 
Using these definitions, the mixture dynamics described by Eq.~\eqref{eq:phidot} can be written in dimensionless form:
\begin{align}
    \partial_{\widetilde t}\phi_i
    &=
    \widetilde{\nabla}\cdot
    \left[
        \frac{D_i}{D_0}
        \phi_i\phi_S
        \widetilde{\nabla}
        \frac{\mu_i}{k_\mathrm{B}T}
        +
        \widetilde{\boldsymbol{\xi}}_i
    \right]
    +
    \frac{L_0^2k_c}{D_0}\tilde{r}_i
    +
    \widetilde{\eta}_i \, .
    \label{eq:dimensionless_phi_dynamics_general_noise}
\end{align}
Here, $\tilde{r}_i=r_i/k_c$ is the dimensionless deterministic reaction term. 

The dimensionless conserved noise $\widetilde{\boldsymbol{\xi}}_i$ and the non-conserved reaction noise $\widetilde{\eta}_i$ have zero mean,
\begin{align}
    \left\langle
        \widetilde{\xi}_{i,\alpha}
        (\widetilde{\mathbf r},\widetilde t)
    \right\rangle
    &=
    0 \, ,
    &
    \left\langle
        \widetilde{\eta}_i
        (\widetilde{\mathbf r},\widetilde t)
    \right\rangle
    &=
    0 \, ,
\end{align}
where $\alpha$ denotes a Cartesian component.
Moreover, both noise terms are white without correlations.
We consider that the fluctuation--dissipation theorem holds. 
Thus, the covariance of the conserved noise reads
\begin{align}
    &\left\langle
        \widetilde{\xi}_{i,\alpha}
        (\widetilde{\mathbf r},\widetilde t)
        \widetilde{\xi}_{j,\beta}
        (\widetilde{\mathbf r}',\widetilde t')
    \right\rangle
    \nonumber\\
    &\qquad =
    2\frac{D_i}{D_0}
    \frac{\nu_i}{L_0^d}
    \phi_i\phi_S\,
    \delta_{ij}\delta_{\alpha\beta}
    \delta^d
    \left(
        \widetilde{\mathbf r}
        -
        \widetilde{\mathbf r}'
    \right)
    \delta
    \left(
        \widetilde t-\widetilde t'
    \right) \, ,
    \label{eq:dimensionless_conserved_noise_covariance}
\end{align}
and the covariance of the non-conserved reaction noise is
\begin{align}
    &\left\langle
        \widetilde{\eta}_i
        (\widetilde{\mathbf r},\widetilde t)
        \widetilde{\eta}_j
        (\widetilde{\mathbf r}',\widetilde t')
    \right\rangle
    \nonumber\\
    &\qquad =
    2\frac{\nu_i}{L_0^d}
    \frac{L_0^2k_c}{D_0}
    \left|\tilde{r}_i\right|\,
    \delta_{ij}
    \delta^d
    \left(
        \widetilde{\mathbf r}
        -
        \widetilde{\mathbf r}'
    \right)
    \delta
    \left(
        \widetilde t-\widetilde t'
    \right) \, .
    \label{eq:dimensionless_reaction_noise_covariance}
\end{align}
 The factor
\begin{align}
    \sigma_i^2
    =
    \frac{\nu_i}{L_0^d}
\end{align}
sets the relative strength of fluctuations. 
It compares the microscopic scale $\nu_i$ with the characteristic scale $L_0^d$ ($d$ is spatial dimension), so fluctuations become weaker for larger systems.\\[0.5em]
\textbf{(i) Perturbing interactions.}
For interaction quenches, the system is driven by changing the interaction parameter $\chi$ in time. 
The relative quench rate is defined as
\begin{align}
    k
    =
    \frac{1}{\Delta\chi}
    \frac{\mathrm d\chi}{\mathrm dt} \, .
    \label{eq:relative_quench_rate}
\end{align}
With the diffusive time scale introduced above, the corresponding dimensionless quench rate is
\begin{align}
    \tilde k
    =
    k\frac{L_0^2}{D_0} \, .
    \label{eq:dimensionless_quench_rate}
\end{align}
For these simulations, we use a square box of side length $L=1.5L_0$, discretized on a $200\times200$ spatial grid, with initial droplet radius $R_0/L = 0.267$.
The diffusion coefficients are reported in units of $D_0$, with
$D_A^\text{I}/D_0=25$ and $D_A^\text{II}/D_0=0.1$.\\[0.5em]
\textbf{(ii) Perturbing composition by reactions.}
For reaction-driven simulations, the same rescaling is used, but the driving enters through the local reaction term. 
For a reaction term of the form
\begin{align}
    r_i = k_c R_i(\{\phi_j\}) \, ,
\end{align}
where $R_i$ is dimensionless, the dimensionless reaction rate is
\begin{align}
    \tilde k_c
    =
    \frac{k_cL_0^2}{D_0} \, .
    \label{eq:reaction_driven_dimensionless_rate}
\end{align}
For the ternary reaction simulations, we use a square box of side length $L=0.005\, L_0$, discretized on a $150\times150$ spatial grid, with dilute-phase diffusion coefficients $D_i^\RN{1}/D_0=1$.

\begin{table*}[tb]
\centering
\small
\setlength{\tabcolsep}{4pt}
\renewcommand{\arraystretch}{1.25}
\begin{tabular*}{\textwidth}{@{\extracolsep{\fill}}lcccccccc}
\hline
Figure 
& $\chi_{AS}$ 
& $\chi_{BS}$ 
& $\chi_{AB}$ 
& $\kappa_A/(k_\text{B}T\,L^2)$ 
& $\kappa_B/(k_\text{B}T\,L^2)$ 
& $D_A^\text{II}/D_0$ 
& $D_B^\text{II}/D_0$ 
& comment\\
\hline\\[-1.25em]
\ref{fig_binary_temperature_quench} 
& 1.65 
& -- 
& -- 
& 0.025 
& -- 
&  0.1 
& -- 
&$\sigma=10^{-5}$, $N=2$\\[0.5em]
\ref{fig3:reactions_kin}(c,g) 
& 3.0 
& 0.6 
& $-0.6$ 
& $4.44\times10^{-3}$ 
& $4.44\times10^{-3}$ 
&  $5\times10^{-1}$
& $10^{-4}$ 
& prod. out/in \\[0.5em]
\ref{fig3:reactions_kin}(d,f) 
& 3.0 
& 0.6 
& $-0.6$ 
& $4.44\times10^{-3}$ 
& $4.44\times10^{-3}$ 
& $10^{-4}$
& $5\times10^{-1}$
& deg. out/in \\[0.5em]
\ref{fig:4asym_diffusion} 
& 3.0 
& 0.6 
& $-0.6$ 
& $4.44\times10^{-3}$ 
& $4.44\times10^{-3}$ 
&  $5\times10^{-1}$ 
& -- 
&  prod. out \\[0.5em]
\ref{fig_app:Stocastic_runs} 
& 1.65 
& -- 
& -- 
& 0.025 
& -- 
&  0.1 
& -- 
& $\sigma=10^{-5}$, $N=2$\\[0.5em]
\ref{fig_binary_chemical_reaction} 
&1.75 
& --
& -- 
& $4.44\times10^{-3}$  
& -- 
& -- 
& -- 
& $N = 2$, $L = 0.05L_0$ \\[0.5em]
\ref{fig_app:other_phase_diagram}(a), \ref{fig:extra_state_diagram}(a)(i, iii)
& $2.01$ 
& $3.0$ 
& $0.0$ 
& $4.44\times10^{-3}$ 
& $4.44\times10^{-3}$ 
& $10^{-4}$ 
& $5\times10^{-1}$ 
& prod. out/in \\[0.5em]
\ref{fig_app:other_phase_diagram}(a), \ref{fig:extra_state_diagram}(a)(ii, iv)
& $2.01$ 
& $3.0$ 
& $0.0$ 
& $4.44\times10^{-3}$ 
& $4.44\times10^{-3}$ 
& $10^{-4}$ 
& $5\times10^{-1}$ 
&deg. out/in \\[0.5em]
\ref{fig_app:other_phase_diagram}(b), \ref{fig:extra_state_diagram}(b)
& $1.79$ 
& $1.79$ 
& $-1.79$ 
& $4.44\times10^{-3}$ 
& $4.44\times10^{-3}$ 
&  $10^{-4}$ 
& $5\times10^{-1}$ 
& -- \\[0.5em]
\ref{fig:state_diagram_with_ic}
& $1.79$ 
& $1.79$ 
& $-1.79$ 
& $4.44\times10^{-3}$ 
& $4.44\times10^{-3}$ 
&  $10^{-4}$ 
& $5\times10^{-1}$ 
& -- \\
\hline
\end{tabular*}
\caption{\textbf{Simulation parameters used in different figures.} 
The internal energies for all components are $\omega_i =0$ in all figures.
The interaction parameters $\chi_{ij}$ in the table are measured in units of $k_\text{B}T$. 
Unless specified otherwise, all components are taken to have equal molecular volumes, corresponding to $N=1$, and their diffusivities in the dilute phase are set to $D_i^\text{I}/D_0=1$.}
\label{tab:simulation_parameters}
\end{table*}

\section{Closed-form binodals and spinodals}\label{appendix:analytics}

Following the procedure in Ref.~\cite{qian2022analytical}, we can analytically approximate binodal and spinodal compositions. This procedure yields a critical interaction strength 
\begin{equation}
    \chi_* = \frac{1}{2}\left(1+\frac{1}{\sqrt{N}}\right)^2\,.
\end{equation}
Here, $\chi_*$ denotes the critical interaction strength above which the mixture can phase separate into distinct dilute $\phi^\text{I}$ and dense $\phi^\text{II}$ phases.
At an interaction strength $\chi_0$ close to this critical interaction strength $\chi_*$, the dense phase composition is approximately
\begin{equation}
    \phi^{\RN{1}/\RN{2}}(\chi_0) \simeq \frac{1}{1+\sqrt{N}} \mp \sqrt{\frac{3\left(\chi_0 - \chi_*\right)}{2\chi_*^2\sqrt{N}}} \,,\label{eq:phiclosetocritpoint}
\end{equation}
where `$-$' is for dilute phase $\RN{1}$ and `$+$' is for dense phase $\RN{2}$.

From the spinodal condition $f''=0$, we find the interaction strength at which a composition enters the spinodal:
\begin{equation}
    \chi_\text{spin}(\phi) = \frac{1}{2}\left(\frac{1}{N\phi_\text{spin}}  + \frac{1}{1-\phi_\text{spin}} \right)\,. \label{eq:chi_spino}
\end{equation}
Thus, the minimal quench depth for vacuoles to form in the dense phase is set by
\begin{equation}
    \Delta\chi_\text{min} = \chi_\text{spin} - \chi_0\, .
\end{equation}
For systems which are perturbed from a state close to the critical point, the approximate composition of the dense phase from Eq.~\eqref{eq:phiclosetocritpoint} can be used to find a closed form expression for the minimal quench depth on the form:
\begin{align}
\begin{split}
    \Delta\chi_{\rm min} &= \frac{1/2}{N\left(\dfrac{1}{1+\sqrt{N}}+\sqrt{\dfrac{3(\chi_0-\chi_*)}{2\chi_*^2\sqrt{N}}}\right)} \\[0.25em] &+ \frac{1/2}{\left(1-\dfrac{1}{1+\sqrt{N}}-\sqrt{\dfrac{3(\chi_0-\chi_*)}{2\chi_*^2\sqrt{N}}}\right)} - \chi_0\,. \label{eq:minimumquench}
\end{split}
\end{align}
Close to the critical point, this expression simplifies to 
\begin{equation}
    \Delta\chi_\text{min}(\chi_0\gtrsim \chi_*) = 2(\chi_0-\chi_*)\,. 
\end{equation}\par 
Far away from the critical point, an alternative expression for the compositions can be used instead:
\begin{align}
    \phi^\RN{2}(\chi) &\simeq 1-\exp{-\gamma_N-\chi}, \label{eq:phi+}\\
    \phi^\RN{1}(\chi) &\simeq \exp{-1-N(\chi-1)} \label{eq:phi-}\,,
\end{align}
meaning it is a more accurate approximation than Eq.~\eqref{eq:phiclosetocritpoint} for cases where $\chi\gg\chi_*$. 
Here, $\gamma_N=1-1/N$. We find a critical quenching depth
\begin{equation}  \Delta\chi_\text{min}(\chi_0\gg\chi_*) \simeq \frac{1}{2}\left(e^{1-1/N+\chi_0} + \frac{1}{N} \right) - \chi_0\,,
\end{equation}
from which we can conclude that vacuoles form when the quench depth is much smaller close to the critical point; see also graphically in Fig.~\ref{fig:2}(c).

\subsection{Critical quench rate for vacuole formation}

To find how the minimal quenching depth changes with composition, we assume that the droplet composition at the center $\phi_\text{c}(t)$ can be described by a linear relaxation dynamics. 
The time-scale of this relaxation (slowest mode) is the diffusive time-scale $R^2/D$ of diffusion on the length scale of the droplet radius $R$:
\begin{equation}
    \dv{\phi_\text{c}}{t} = -\frac{D}{R^2}\left(\phi_\text{c}-\phi_\text{b}(t)\right)\,,
\end{equation}
where $D$ is the diffusion coefficient in the dense phase, $\phi_\text{b}(t)$ is the dynamic composition of the droplet at the droplet boundary moving along the dense binodal branch. 
Moreover, for simplicity, we approximate the droplet radius $R$ as a constant and will later evaluate it using the final radius. 
The composition of the droplet at its boundary is approximated to change linearly in time according to
\begin{equation}
    \phi_\text{b}(t) = \phi^\text{II}(\chi_0) + vt\,,\qquad 0\leq t \leq \tau_q\,,
\end{equation}
where $\phi^\text{II}(\chi_0)\equiv\,\phi^\text{II}_0$ is the initial composition at equilibrium with an interaction strength $\chi_0$ at $t=0$, and $v$ is the compositional velocity at the droplet boundary $v=k(\phi^\text{II}(\chi_\infty) - \phi^\text{II}(\chi_0))$ such that the final equilibrium composition is $\phi^\text{II}(\chi_\infty)\equiv\,\phi^\text{II}_\infty$ at $t\rightarrow\infty$. 
We define the deviation from the center composition to the boundary composition as $\Lambda=\phi_\text{b}-\phi_\text{c}$, which has a solution
\begin{equation}
    \Lambda(t) = \frac{kR^2(\phi^\text{II}_\infty - \phi_0^\text{II})}{D}\left(1-\exp{-Dt/R^2}\right)\,.
\end{equation}
At the final time-point of the quench $\tau_\text{q}$ the deviation is
\begin{equation}
    \Lambda(\tau_\text{q}) = \frac{kR^2(\phi^\text{II}_\infty - \phi_0^\text{II})}{D}\left(1-\exp{-\frac{D}{kR^2}}\right)\,.
\end{equation}
A vacuole forms when the composition enters the spinodal, where $f''=0$, i.e. $\Lambda(\tau_\text{q})\geq\phi^\RN{2}_\infty-\phi^\RN{2}_\text{spin}(\chi_\infty)$, where $\phi^\RN{2}_\infty$ and $\phi^\text{II}_\text{spin}$ are the dense and spinodal composition at the final interaction strength, yielding
\begin{equation}
     \frac{kR^2}{D}\left(1-\exp{-\frac{D}{kR^2}}\right)=\frac{\phi^\RN{2}_\infty-\phi^\text{II}_\text{spin}(\chi_\infty)}{\phi^\text{II}_\infty - \phi^\text{II}_0}\,. \label{eq:trancendental}
\end{equation}
We note that the left-hand-side is always less than $1$ of the equation above, such that finding vacuoles requires $\phi_\text{spin}^\RN{2}(\chi_\infty)>\phi^\RN{2}_0$, which can be graphically understood from Fig.~\ref{fig_binary_temperature_quench}. 
The equation is a transcendental equation for $kR^2/D$, which can be solved in a closed form by performing a Taylor expansion to second order in $D/(kR^2)$.
Finally, we find Eq.~\eqref{eq:threshold} describing the minimal quench rate for vacuole formation.
This expression ($\phi^\RN{2}_\infty$ is evaluated through Eq.~\eqref{eq:phi+}, $\phi^\text{II}_0$ through Eq.~\eqref{eq:phiclosetocritpoint}, and $\phi^\text{II}_\text{spin}$)
can be found from inverting Eq.~\eqref{eq:chi_spino}:
\begin{equation}
    \phi^\RN{2}_\text{spin}(\chi) = \frac{1}{2}-\frac{1-\frac{1}{N}}{4\chi} + \sqrt{\left(\frac{1}{2}- \frac{1-\frac{1}{N}}{4\chi}\right)^2 - \frac{1}{2\chi N} }\,.
\end{equation}
Note that since the LHS of Eq.~\eqref{eq:trancendental} is always smaller or equal to $1$, we can obtain the minimal quench depth by setting the RHS equal to one. This yields the same expression as in Eq.~\eqref{eq:minimumquench}.

\subsection{Maximal vacuole volume}

From mass conservation, we determined a lever-rule for the maximal possible vacuole size relative to the droplet. The maximal size is then set by thermodynamic parameters:
\begin{equation}
    \text{max}_t\left(\frac{A_\text{vac}(t)}{A_\text{drop}}\right) \simeq \frac{\phi^\RN{2}(\chi_0)-\phi^\RN{1}(\chi_\infty)}{\phi^\RN{2}(\chi_\infty)-\phi^\RN{1}(\chi_\infty)}\,, \label{eq:max_Vvac}
\end{equation}
Note that the lever-rule is independent of dimension, resulting in the same expression in $d=2$ as in $d=3$. 
As our simulations are two-dimensional ($d=2$), we write the expression using area $A$ instead of volume $V$.
Using the expression for the dense and dilute binodal composition $\phi^\text{II/I}(\chi_\infty)$ in Eq.~\eqref{eq:phi+} and Eq.~\eqref{eq:phi-}, respectively, and composition of droplet before the quench is $\phi^\RN{2}(\chi_0)$ in Eq.~\eqref{eq:phiclosetocritpoint}.
This expression yields a completely analytic expression for the relative size, as given in Eq.~\eqref{eq:max_vac_area}. 
For long polymers ($N\gg 1$), the expression can be expanded:
\begin{equation}
    \text{max}_t\left(\frac{A_\text{vac}(t)}{A_\text{drop}}\right) \simeq 1-\frac{1}{1-e^{-1-\chi_\infty}}\sqrt{\frac{3(\chi_0-\chi_*)}{2\chi_*^2}}N^{-1/4}\,.
\end{equation}
This limit shows that large vacuoles are  favored for condensates scaffolded by long polymers ($N\gg1$).

\section{Effects of thermal fluctuations on vacuole formation}
\label{sec:app:thermal_fluctuation}

\begin{figure*}[tb]
    \centering
    \includegraphics[width=\textwidth]{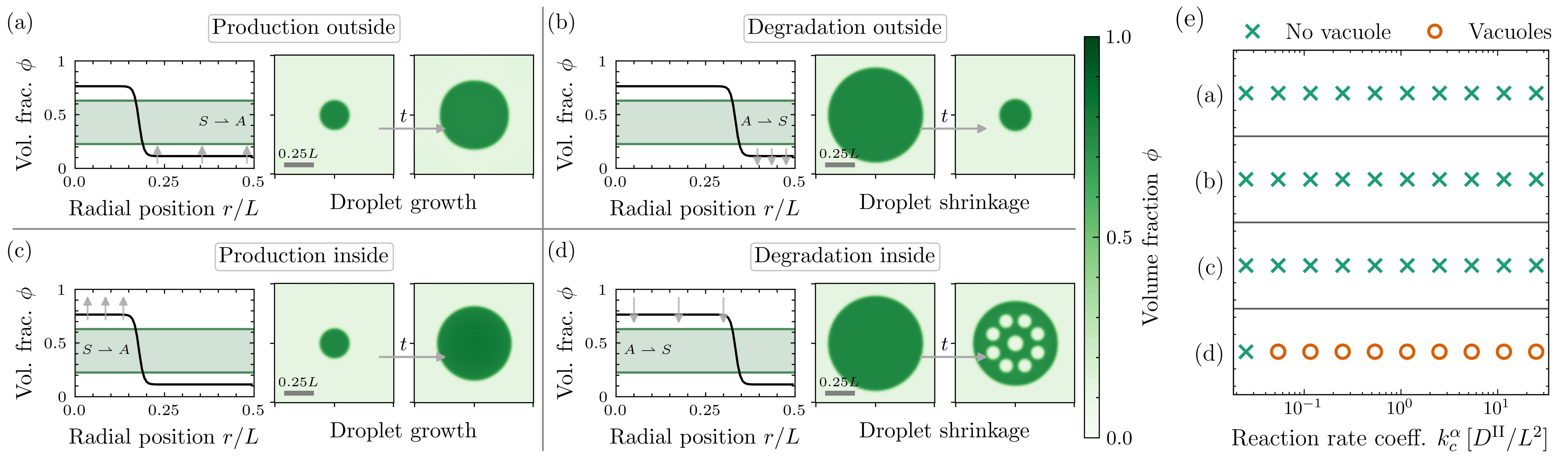}
    \caption{\textbf{Phase separating binary mixture undergoing chemical reactions can form vacuoles:}
    (a-d) A binary mixture that phase-separates to form a droplet undergoes (production $S \rightharpoonup A$/degradation $A \rightharpoonup S$) chemical reactions inside or outside the droplet.
    These reactions affect the composition locally as indicated by grey arrows.
    Production reactions lead to droplet growth, and the degradation reactions shrink the droplet. 
    (e) The formation of vacuoles is promoted when a degradation reaction is triggered inside the droplet.}
    \label{fig_binary_chemical_reaction}
\end{figure*}

\begin{figure}[h!]
    \centering
    \includegraphics[width=0.75\linewidth]{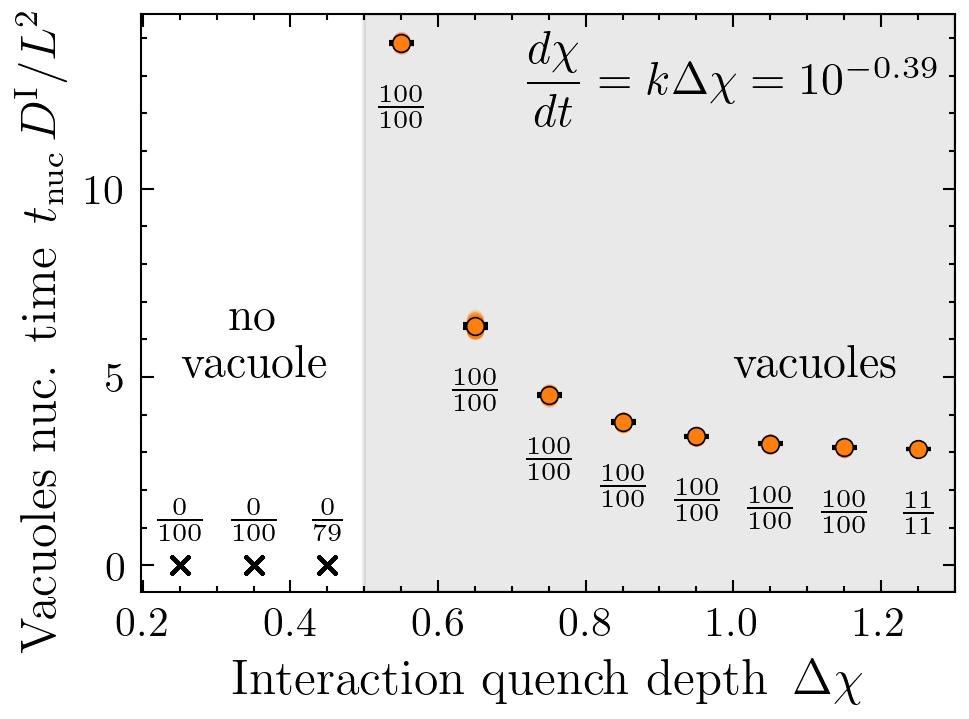}
    \caption{
    \textbf{Vacuole nucleation is robust to thermal fluctuations:}
    Below a critical quench depth ($\Delta\chi< \Delta\chi_\text{min}$, white domain), the droplet relaxes without forming a vacuole, whereas above this threshold ($\Delta\chi> \Delta\chi_\text{min}$, gray domain) vacuoles nucleate reproducibly. 
    The small spread in nucleation times indicates that thermal fluctuations do not noticeably alter the vacuole-formation threshold.}
    \label{fig_app:Stocastic_runs}
\end{figure}

The vacuoles formed through perturbing interactions are dominantly controlled by the deterministic driving term in Eq.~\eqref{eq:phidot}. 
To test whether thermal fluctuations can modify this behavior, we performed stochastic simulations starting from an initially phase-separated droplet. 
The system was then quenched by changing the molecular interaction parameter $\chi$ at a fixed interaction velocity, ${d\chi}/{dt}=k\Delta\chi$.
We record the time at which a vacuole nucleates inside the droplet for different realizations. 
Fig.~\ref{fig_app:Stocastic_runs} shows such simulation results for different values of the quench depth $\Delta\chi$, with the fraction shown above the marker denoting the number of stochastic realizations in which a vacuole formed. 
These simulations show that vacuoles form only when the quench depth exceeds a critical value $\Delta\chi_\text{min}$. 
Below this threshold, no vacuoles are observed in the stochastic simulations, whereas above it, vacuoles form reproducibly across all tested noise realizations. 
Thus, for the noise strength considered here, vacuole formation is mainly controlled by the deterministic drive. 
Thermal fluctuations neither noticeably shift the critical quench nor cause significant variations in the vacuole nucleation time. However, fluctuations can affect where vacuoles form. 

For vacuole formation driven by chemical processes, we find that thermal fluctuations do not significantly affect the overall state diagram; see Fig.~\ref{fig3:reactions_kin}. 
This is similar to the interaction-quench case, where fluctuations primarily influence the spatial location where vacuoles nucleate, while the timing of vacuole formation is only weakly affected.
Therefore, for computational efficiency, the simulations shown in Fig.~\ref{fig3:reactions_kin} and Fig.~\ref{fig:4asym_diffusion} were performed without  fluctuations.

\begin{figure}[h]
    \centering
    \includegraphics[width=0.675\linewidth]{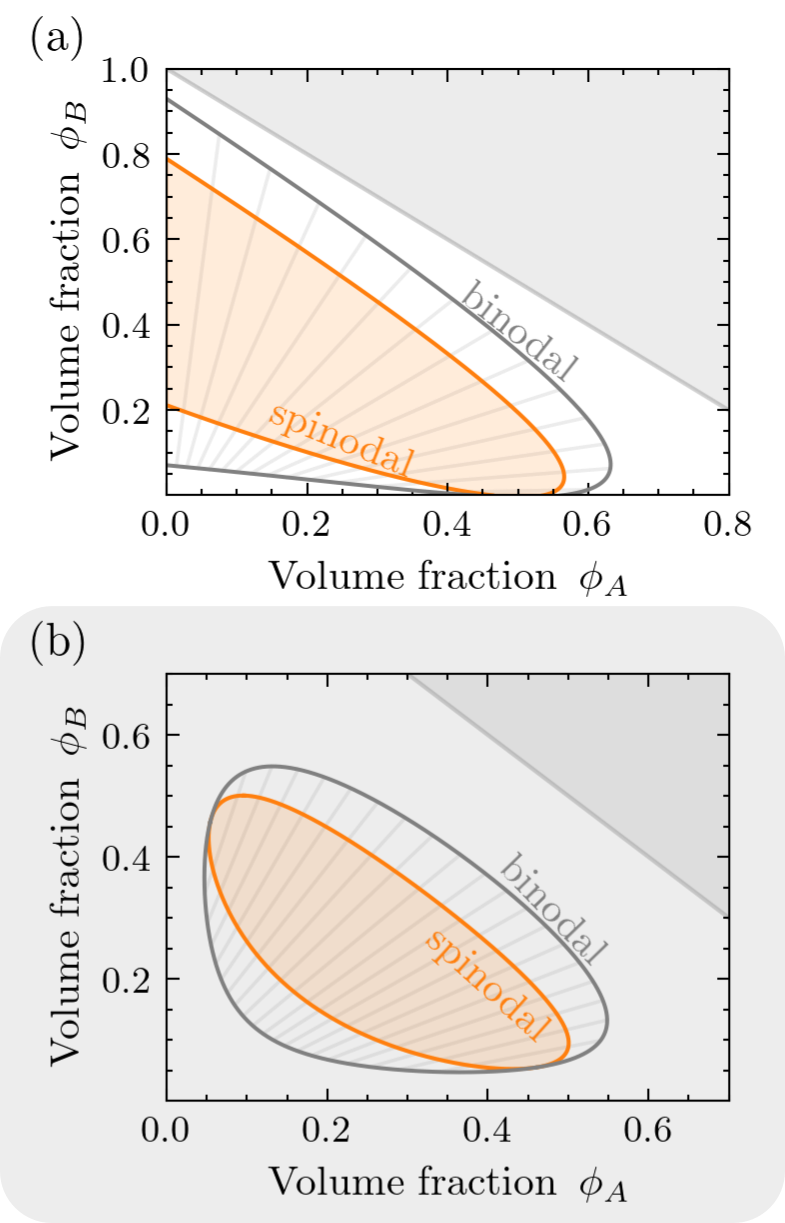}
    \caption{\textbf{Ternary phase diagrams for different interaction classes.}
    Representative phase diagrams for two multicomponent liquid mixtures used to test the robustness of reaction-driven vacuole formation. 
    (a) Both $A$ and $B$ are solvent-phobic, while $A$ and $B$ attract each other. 
    (b) Both $A$ and $B$ are solvent-philic, while $A$ and $B$ attract each other.}
    \label{fig_app:other_phase_diagram}
\end{figure}

\section{Vacuole formation by chemical reactions in binary systems}
\label{sec:app:binary_react}

\begin{figure*}[tb]
    \centering
    \makebox[\textwidth][c]{%
        \hspace*{-0.03\textwidth}%
        \includegraphics[width=1.035\textwidth]{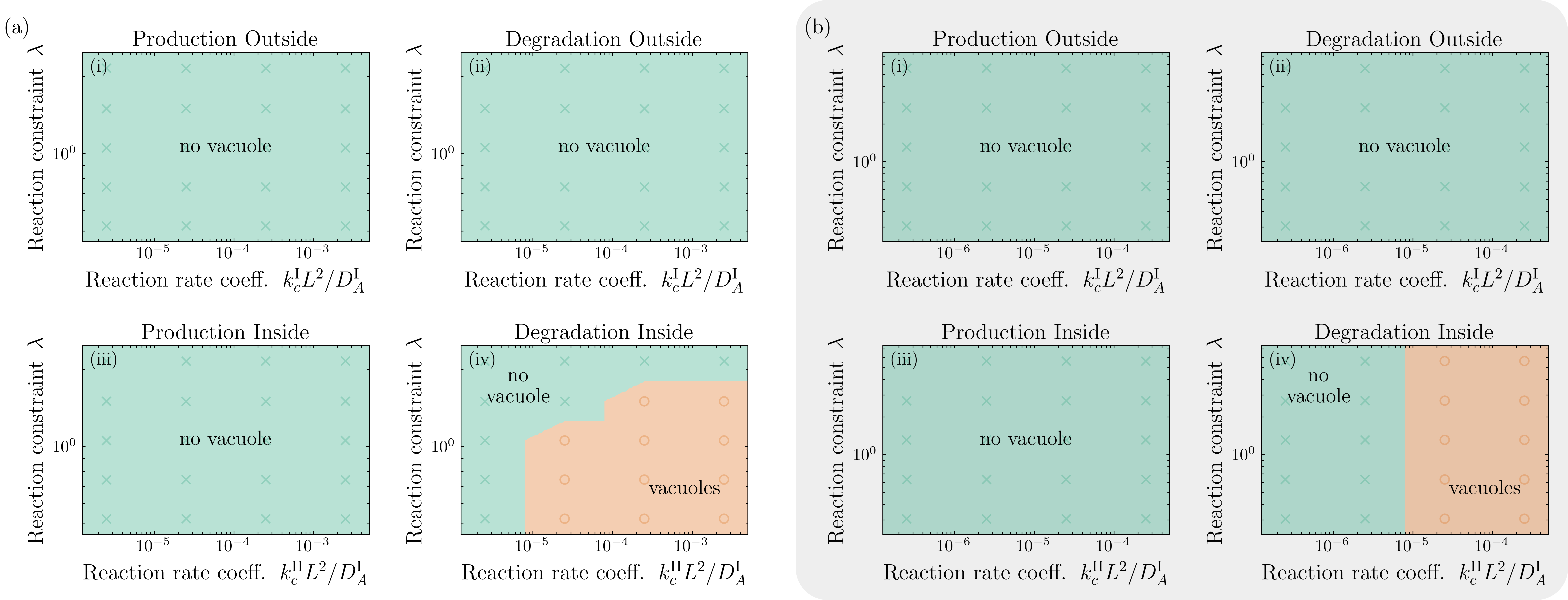}
    }
    \caption{
    \textbf{Vacuole formation in multicomponent liquid mixtures with different interactions.}
    State diagrams characterizing vacuole formation as a function of the reaction constraint $\lambda$ and the  reaction rate coefficient $k_c L^2/D_A^I$. 
    (a) corresponds to the interaction class in which $A$ and $B$ are solvent-phobic, while $A$ and $B$ attract each other. 
    (b) corresponds to the interaction class in which $A$ and $B$ are solvent-philic, while $A$ and $B$ attract each other (see Fig.~\ref{fig_app:other_phase_diagram}(a,b) for phase-diagram). 
    Crosses denote simulations in which no vacuole forms, whereas circles denote simulations in which vacuoles form. 
    For the interaction classes considered here, vacuole formation is observed only when degradation occurs inside the droplet and only at sufficiently high reaction rates.}
    \label{fig:extra_state_diagram}
\end{figure*}
In a binary system, we investigate two simple unidirectional reactions between $S$ and $A$:
\begin{align}
    A &\rightharpoonup S:\quad \,\,r_A = -k_A(\phi_S)\,\,\phi_A,\,\\ S &\rightharpoonup A:\quad \,\,r_A = k_S(\phi_S)\,\,\phi_S\,,
\end{align}
and $r_S=-r_A$, where $\phi_S=1-\phi_A$. The reaction rate coefficient $k_i$ depends on $\phi$, such that the reaction rate is different between the phases (see Eq.~\eqref{eq:k_c_switch} for the complete expression). 
We will investigate these two reactions in two different cases, where the reaction either occurs in the dense phase or the dilute phase.

Out of the four cases (production (a,c)/degradation (b,d), inside(c,d)/outside(a,b) in Fig.~\ref{fig_binary_chemical_reaction}), only degradation inside produces vacuoles. 
The droplet grows for the production reaction ($S\rightharpoonup A$) and shrinks under degradation ($A\rightharpoonup S$). 
Because the equilibrium dense phase composition cannot change as the reaction proceeds, degradation inside the droplet is the only case in which the dense-phase composition moves toward the spinodal, and hence the only one that produces vacuoles.
These gradients in composition touches the spinodal, allowing for local instablity of the droplet phase leading to the formation of vacuoles.
For vacuoles to nucleate, the reaction rate coefficient must be sufficiently fast compared to the diffusion time-scale over the droplet size, as seen in Fig.~\ref{fig_binary_chemical_reaction}(e).

\section{Vacuole formation in multi-component systems with different binodals}
\label{sec:app:vary_bino}

\begin{figure*}[tb!]
    \centering
    \includegraphics[width=0.9\textwidth]{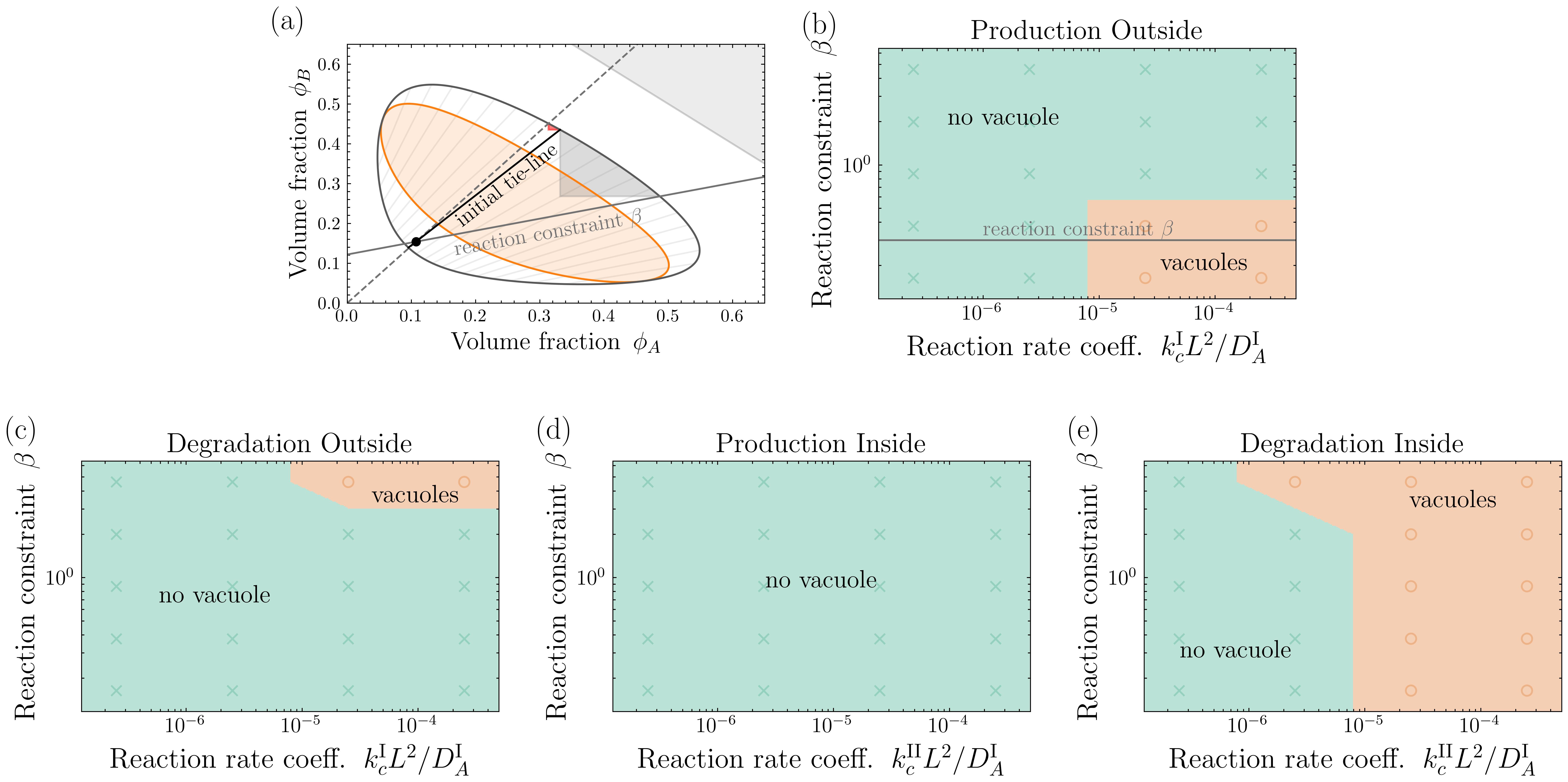}
    \caption{
    \textbf{Vacuole formation is enhanced when reaction trajectories deviate from tie-lines.}
    (a) Ternary phase diagram for the solvent-philic interaction class. 
    For the original initial condition, the reaction-constraint line $\lambda$ is nearly parallel to the initial tie-line, so that the accessible composition domain, shown in red, remains narrow and outside the spinodal region. 
    In this case, reaction-induced composition gradients do not trigger a local instability inside the droplet. 
    Initializing the chemical perturbation with a different reaction constraint $\beta$ (shown in grey line) enlarges the accessible composition domain, shown in grey. 
    This allows the droplet composition to enter the spinodal region. 
    (b--d) State diagrams for modified chemical perturbation, showing that vacuoles can form for the same interaction class when reaction-driven composition changes access the unstable region of the phase diagram.}
    \label{fig:state_diagram_with_ic}
\end{figure*}

To understand the role of liquid-mixture interactions in reaction-driven vacuole formation, we compare additional interaction classes using the same chemical reaction scheme to the one in the main text. 
For simplicity, we consider two different types of interactions. 
In the first case (see Fig.~\ref{fig_app:other_phase_diagram}(a)), both $A$ and $B$ are solvent-phobic, while $A$ and $B$ attract each other, such that the two components tend to phase separate from the solvent-rich phase. 
In the second case (see Fig.~\ref{fig_app:other_phase_diagram}(b)), both $A$ and $B$ are solvent-philic, while $A$ and $B$ also attract each other. 
For each interaction class, we summarize the results in four state diagrams, as shown in Fig.~\ref{fig:extra_state_diagram}(a,b).
We find that most reaction schemes do not lead to vacuole formation.
Production inside or outside the droplet leads to droplet growth, whereas degradation outside mainly causes droplet shrinkage, with neither case producing vacuoles.
In contrast, when degradation occurs inside the droplet, vacuole formation is observed for sufficiently fast chemical reactions.
To understand why vacuole formation is not pronounced for such systems, consider the hydrophilic interacting  phase diagram as shown in Fig.~\ref{fig:state_diagram_with_ic}(a). 
We observe that as the reaction-constraint line $\lambda$ (shown in dashed line) becomes parallel with the initial tie-line. This makes the behavior similar to the binary system, where only degradation in the dense phase nucleated vacuoles, as the equilibrium dense phase composition remains close to unchanged during the reaction.
Consequently, the triangular domain accessible to the composition trajectories, shown in red, narrows down and does not enter the spinodal region; hence, no vacuole is formed.

The formation of a vacuole, however, may become much more pronounced when we start with a different initial condition. 
For example, we consider the initial tie-line shown in black in Fig.~\ref{fig:state_diagram_with_ic}(a).
For production, this initial tie is selected such that the droplet occupies $5\%$ of the system volume; then the chemical reactions in the system occur such that the production of $B$ to $A$ always maintains a ratio $\beta$.
Choosing a ratio $\beta$ that differs from $\lambda$ makes the reaction-constraint line non-parallel to the initial tie-line. 
This enlarges the accessible triangular domain, shown in grey, allowing it to extend deeper into the spinodal region. 
Consequently, the droplet trajectories can enter the spinodal region, making vacuole formation more pronounced.
We summarize these results in Fig.~\ref{fig:state_diagram_with_ic}(b-e), where vacuole formation is now observed for the same interaction class.

%
\end{document}